\documentclass{aa}  

\usepackage{graphicx}
\usepackage[varg]{txfonts}

\usepackage{hyperref}
\usepackage{natbib}
\makeatletter 
\renewcommand*\aa@pageof{, page \thepage{} of \pageref*{LastPage}}
\makeatother

\begin{document}

\title{Examining the detectability of ringing on highly eccentric exoplanets}

\author{
    M. Vanrespaille\inst{1}
    \and
    R. Baeyens\inst{2}\fnmsep
    \and
    A. Schneider\inst{1,3}
    \and
    L. Carone\inst{4}
    \and
    L. Decin\inst{1}
}

\institute{
    Institute for Astronomy (IvS), KU Leuven, Celestijnenlaan 200D, 3001 Leuven, Belgium\\
    \email{mathijs.vanrespaille@kuleuven.be} 
    \and
    Anton Pannekoek Institute for Astronomy, University of Amsterdam, Science Park 904, 1098 XH Amsterdam, The Netherlands
    \and
    Centre for ExoLife Sciences, Niels Bohr Institute, Øster Voldgade 5, 1350 Copenhagen, Denmark
    \and
    Space Research Institute, Austrian Academy of Sciences, Schmiedlstrasse 6, 8042 Graz, Austria
}

   \date{Received September 15, 1996; accepted March 16, 1997}

\abstract
   {Eccentric exoplanets offer an opportunity to study the response of an atmosphere to changing thermal forcing and the robustness of the super-rotating equatorial jet seen on tidally locked hot Jupiters. However, the atmospheric dynamics on eccentric planets strongly depend on the planetary rotation period, which is difficult to constrain observationally. The ringing phenomenon, whereby the observed emission increases and decreases after the periastron passage as the flash-heated hemisphere rotates into and out of view, can provide a tight constraint on rotation. }
   {We studied five highly eccentric transiting exoplanets \mbox{\object{HAT-P-2 b}}, \mbox{\object{HD 80606 b}}, \mbox{\object{TOI-3362 b}}, \mbox{\object{TOI-4127 b}} and \mbox{\object{HD 17156 b}} to find which displays strong ringing signals that are sufficiently strong for the \textit{James Webb} Space Telescope (JWST) to detect.}
   {We implemented the treatment of eccentricity and non-synchronous rotation in the non-grey climate model \texttt{expeRT/MITgcm} and generated synthetic light curves. }
   {We find four detectable ringing peaks on \mbox{\object{HD 80606 b}} and some undetectable ringing on \mbox{\object{TOI-4127 b}} and \mbox{\object{HD 17156 b}}. The lack of clouds, photo-chemistry and obliquity in our models may have led us to overestimate the amplitude of the ringing however. The strength of the ringing signal is mostly determined by the eccentricity, planetary rotation period, planet-to-star radius ratio and apparent magnitude of the system. We searched for more exoplanets that could show ringing but found no candidates as promising as \mbox{\object{HD 80606 b}}. }
   {We recommend prioritising \mbox{\object{HD 80606 b}} as a target for ringing with JWST. A baseline of five days after the periastron passage would capture three ringing peaks, which is sufficient to tightly constrain the planetary rotation period. An extension to seven days would add a fourth peak, which would allow us to verify the rotation period. }

   
   \maketitle

\section{Introduction}

In the past few decades there have been great advances in the study of exoplanetary atmospheres. The observations of the recently launched \textit{James Webb }Space Telescope (JWST) have already contributed greatly to these efforts \citep[e.g.][]{JWST2023,Feinstein2023,Tsai2023b,Dyrek2023}. Because of the complicated nature of exoplanetary atmospheres, observations should be compared to global circulation models (GCMs) to be interpreted properly. These GCMs require well-constrained orbital and planetary parameters, such as the planet's radius, rotation period, and orbital eccentricity, as input \citep{Showman2020}. If the planet transits its parent star, most of these quantities can be reliably measured, with the notable exception of the rotation period \citep{Sing2018}. 

Most studies so far have focussed on hot Jupiters because their size and brightness makes it relatively easy to achieve a high observational signal-to-noise \citep{Showman2020}. As a result of their close-in orbits, hot Jupiters are often on a tidally circularised orbit. Based on timescale arguments, it can then safely be assumed that the planet's rotation period is equal to the orbital period \citep{Hut1981,Socrates2012,Rauscher2017} and that its obliquity is zero \citep{Rauscher2017,Rauscher2023}. Consequently, the irradiation these planets receive is constant in intensity and always illuminates the same hemisphere. On the dayside hemisphere, the illumination creates a pattern of geostrophic winds that transport angular momentum towards the equator \citep{Showman2010,Showman2011}. Consequently, these winds form a remarkable circulation pattern that is dominated by an eastward flow of gas centred on the equator and extending around the planet, which typically reaches speeds of several km~s$^{-1}$. This flow is called the equatorial super-rotating jet. The existence of such jets is well established both theoretically \citep[e.g.][]{Showman2002,Showman2009,Tsai2014,Amundsen2016,Carone2020,Showman2020} and observationally \citep[e.g.][]{Knutson2007,Snellen2010,Zellem2014,Louden2015,May2022}. Because of the major role of the Coriolis force in atmospheric dynamics, the planet's rotation period is a decisive parameter in the shaping of these circulation patterns \citep[e.g.][]{Showman2009,Showman2010,Rauscher2014,Showman2015,Baeyens2021a,May2022}. 

Planets on eccentric orbits can no longer be assumed to rotate synchronously \citep{Hut1981}. 
Despite the increased difficulty, studying eccentric and non-synchronously rotating exoplanets is worthwhile as it allows one to examine the response of the atmosphere's dynamics, chemistry, and cloudiness to changing thermal forcing \citep{Kataria2013,Lewis2017,Tsai2023a}. 
Moreover, transit spectroscopy of cool planets on far-out, circular orbits is rarely available as the distance to the star reduces the likelihood that a transit would take place. For an eccentric planet, on the other hand, a transit is more likely to take place around periastron, when the atmosphere still contains traces of the cooler regime around apastron that is otherwise difficult to observe. Finally, about half of the known Jupiter-like planets are eccentric, especially those with a longer orbital period due to the distance dependence of tidal circularisation \citep{Malmberg2009,Dawson2013,Bitsch2020}. 

A number of previous studies have examined the role of eccentricity and non-synchronous rotation in the atmospheric circulation of hot and warm Jupiters. \citet{Showman2009}, \citet{Rauscher2014}, and \citet{Showman2015} used 3D GCMs to study the effect of non-synchronous rotation, concluding that a shorter rotation period weakens the equatorial jet. They disagree on the effect of an increased rotation period, though, with \citet{Rauscher2014} claiming that a westward flow develops and \citet{Showman2009,Showman2015} finding that it leads to significantly broader and faster eastward jets. Nonetheless, these studies all show that the rotation period plays a major role in atmospheric circulation. In a similar vein, \citet{Kataria2013} examined the impact of eccentricity in a theoretical scenario, pointing out that while jets are still ubiquitous, they tend to be faster the greater the eccentricity is and change quite notably in the upper atmosphere throughout the orbit. This is in agreement with the results of \citet{Lewis2014}, who simulated the eccentric planet \mbox{\object{HAT-P-2 b,}} and \citet{Lewis2017} and \citet{Tsai2023a}, who examined the extremely eccentric \mbox{\object{HD 80606 b}}. The latter two studies show that the jet speed on \mbox{\object{HD 80606 b}} decreases between periastron passages. \citet{Lewis2017} also highlight the impact of the rotation period on the circulation, composition, and lofting of clouds and, consequently, the photometric observations of highly eccentric planets. 

During the periastron passage, when an eccentric planet is briefly but intensely heated, most of the flash-heating occurs on the side facing the star. As the planet moves farther from the star and cools, that heated hemisphere rotates into and out of view of an observer on Earth. Consequently, the total brightness of the star and planet combined may display a number of local maxima, which is known as ringing. Thus, by studying the system's light curve after the periastron passage, one can constrain the planet's rotation period as well as how efficiently the temperature is homogenised by circulation. Using a simple thermal model without consistent dynamics or radiative transfer, \citet{Cowan2011} predicted the presence of such ringing on the exoplanets \mbox{\object{HAT-P-2 b}}, \mbox{\object{HD 80606 b,}} and \mbox{\object{HD 17156 b,}} among others. This stands in contrast to the work of \citet{Iro2010}, who, with another simple radiative model, found little ringing in their synthetic light curves of these three planets. \citet{Kataria2013} examined the phenomenon in a theoretical case with a 3D GCM coupled to radiative transfer, highlighting the importance of the viewing geometry. Ringing has not yet been observed on any exoplanet, though. \citet{deWit2016} and \citet{Lewis2017} used the lack of ringing in \textit{Spitzer} observations of \mbox{\object{HD 80606 b}} to constrain its rotation period and infer the likely presence of clouds, demonstrating that even the non-detection of ringing holds valuable information in some systems. As these \textit{Spitzer} observations had a short baseline --  a maximum of two days after the periastron passage and shorter still at the longer wavelengths more suitable to observe ringing -- this does not exclude the possibility that a future JWST observation with a longer baseline may detect ringing. 

No recent work employing a 3D GCM with full coupling to a non-grey radiative transfer implementation has simulated multiple known eccentric exoplanets to predict their photometric signals. Therefore, it is as of yet unclear how many and which planets are likely to display ringing. 
In this study we generated synthetic light curves of five highly eccentric transiting exoplanets to examine whether JWST would be able to detect any ringing. Our ultimate goal is to inform the planning of JWST observing time by identifying viable targets and determining the baseline required for a ringing observation.

\section{Eccentric exoplanet GCM}

\subsection{\texttt{expeRT/MITgcm}}

We employed the atmospheric model \texttt{expeRT/MITgcm}  \citep{Carone2020,Schneider2022a}, which features 3D hydrodynamics coupled to a full non-grey radiative transfer solution. It is built upon the \texttt{MITgcm} dynamical core of \citet{Adcroft2004}, which solves the primitive equations of hydrodynamics assuming an ideal gas. The model also assumes the planet to be a rigidly rotating sphere. We used a C32 cubed-sphere grid with 47 vertical layers, the upper 41 of which are logarithmically spaced in pressure between $10^{-5}$ and 100~bar and the lower six linearly spaced in steps of 100~bar down to 700~bar. A dynamical time-step of 25~s reliably obeys the Courant-Friedrichs-Lewy stability criterion \citep{Courant1928}. To smoothen small-scale noise, a fourth-order Shapiro filter with a 1~day timescale is applied \citep{Shapiro1971}. At the top boundary, unphysical gravity waves are dampened by a soft sponge layer, as described by \citet{Schneider2022a}. Meanwhile, we applied Rayleigh friction below 200 bar to emulate magnetic drag, which \citet{Carone2020} find improves the stability of the model. We also employed the heating term due to this parameterised drag introduced by \citet{Schneider2022b}. 

\texttt{expeRT/MITgcm} features a radiative transfer treatment including scattering using the Feautrier method with accelerated lambda iteration. Following \citet{Schneider2022a}, we used a radiative time-step of 100~s, meaning the radiation field is only updated after every fourth dynamical time-step. Moreover, the radiative flux is only computed for every second column and then interpolated in between. We used six \mbox{correlated-k} bins \citep{Goody1989} with sixteen Gaussian quadrature points each to discretise our 0.26~--~300~$\mu$m wavelength range \citep{Schneider2022a}. We employed the pre-computed opacity table of the chemical equilibrium model of \citet{Molliere2019}, which minimises the Gibbs free energy on a grid of 1000 temperatures at our 47 pressures. It includes: the molecules H$_2$O, CO$_2$, CH$_4$, NH$_3$, CO, H$_2$S, HCN, PH$_3$, and FeH and the atoms Na and K as line opacity sources; Rayleigh scattering by H$_2$ and He; and absorption collisionally induced by interactions between H$_2$-H$_2$, H$_2$-He, and H$^-$. TiO and VO were not added in our models as they often produce temperature inversions in the models \citep[e.g.][]{Showman2009,Lewis2014,Kataria2015,Schneider2022a} and such inversions are only observed on the hottest exoplanets \citep{Heng2018,Arcangeli2018,Parmentier2018,Lothringer2018}. The effects of clouds and hazes were not included in our model. Since measurements of the obliquity are available for few exoplanets and typically have a large uncertainty \citep{Rauscher2017,Ohno2019b,Adams2019,Bryan2020,Bryan2021}, we neglected obliquity in our model. Finally, we assumed each atmosphere has a solar chemical composition, as is commonly done \citep[e.g.][]{Showman2009, Kataria2013, Amundsen2016, Lewis2017}.

\subsection{Implementation of eccentricity and non-synchronous rotation}

We extended \texttt{expeRT/MITgcm} to allow for non-synchronous rotation and eccentric orbits as it was limited to tidally locked planets before. On tidally locked planets, the incoming irradiation is constant in intensity and always centred on the same point. On an eccentric and non-synchronously rotating planet on the other hand, the irradiation varies in two ways. Firstly, the intensity changes as $r^{-2}$, with distance $r$, which depends on the true anomaly $\nu$ as $r = a (1 - e^2) (1 + e \cos{\nu})^{-1}$. Herein, $a$ is the semi-major axis and $e$ the eccentricity. Secondly, the position of the substellar point in the planet's co-rotating reference frame as $\phi_\textrm{ss} = \nu - 2\pi \frac{t}{P_\textrm{rot}}$ with $t$ the time and $P_\textrm{rot}$ the rotation period. 

Since there is no analytical formula to calculate $\nu$ for any given $t$, we applied an iterative Newton-Raphson method to the Kepler equation \citep{Gregory2006} until the error in time was smaller than $10^{-6}$ of the orbital period at each radiative time-step. This method needs no more than ten iterations, making it both precise and computationally cheap. However, the radiative transfer requires more iterations when the temperature structure changes due to the changing irradiation, which can increase the total computation time of the model by a factor of up to two.

\subsection{Setup of the simulations}

\begin{table*}
    \caption[Orbital, planetary and stellar parameters.]{Relevant orbital, planetary, and stellar parameters of the five planetary systems in our sample.}
    \label{tab:planet parameters}
    \centering
    \begin{tabular}{c c c c c c c c c c c c}
        \hline\hline
          & $a$ & $P_\textrm{orb}$ & $e$ & $P_\textrm{ps}$ & $\omega$ & $R_\textrm{p}$ & $M_\textrm{p}$ & $R_*$ & $T_\textrm{eff}$ & m$_\textrm{V}$ & References \\
          & (au) & (d) & (...) & (d) & ($\degr$) & (R$_\textrm{J}$) & (M$_\textrm{J}$) & (R$_\sun$) & (K) & (...) & \\
        \hline
        \mbox{\mbox{\object{HAT-P-2 b}}} & 0.06740 & 5.6334729 & 0.5171 & 1.8927 & 185.2 & 0.951 & 8.74 & 1.54 & 6414 & 8.69 & 1, 2, 3, 4, 5\\
        HD~80606~b & 0.4603 & 111.436765 & 0.9318 & 1.7314 & 301.213 & 1.032 & 4.1641 & 0.98 & 5645 & 9.00 & 5, 6, 7, 8\\
        \mbox{\mbox{\object{TOI-3362 b}}} & 0.153 & 18.09547 & 0.815 & 1.2997 & 51 & 1.142 & 5.03 & 1.830 & 6532 & 10.86 & 5, 9\\
        \mbox{\object{TOI-4127 b}} & 0.3081 & 56.39879 & 0.7471 & 6.6118 & 129.6 & 1.096 & 2.30 & 1.293 & 6096 & 11.44 & 5, 10\\
        HD~17156~b & 0.1623 & 21.216398 & 0.6768 & 3.6775 & 121.71 & 1.095 & 3.191 & 1.508 & 6079 & 8.16 & 5, 11, 12\\
        \hline
    \end{tabular}
    \tablefoot{From left to right, these parameters are the semi-major axis, orbital period, eccentricity, planetary rotation period, argument of periastron, planetary reference radius, planetary mass, planetary surface gravity, stellar radius, stellar effective temperature and apparent $V$-band magnitude. }
    \tablebib{(1)~\citet{Southworth2010}; (2)~\citet{Pal2010}; (3)~\citet{Triaud2010}; (4)~\citet{Tsantaki2014}; (5)~\citet{Hog2000}; (6)~\citet{Pearson2022}; (7)~\citet{Moutou2009}; (8)~\citet{Naef2001}; (9)~\citet{Dong2021}; (10)~\citet{Gupta2023}; (11)~\citet{Winn2009}; (12)~\citet{Nutzman2011}.}
\end{table*}

\begin{figure*}
    \centering
    \includegraphics[width=140mm]{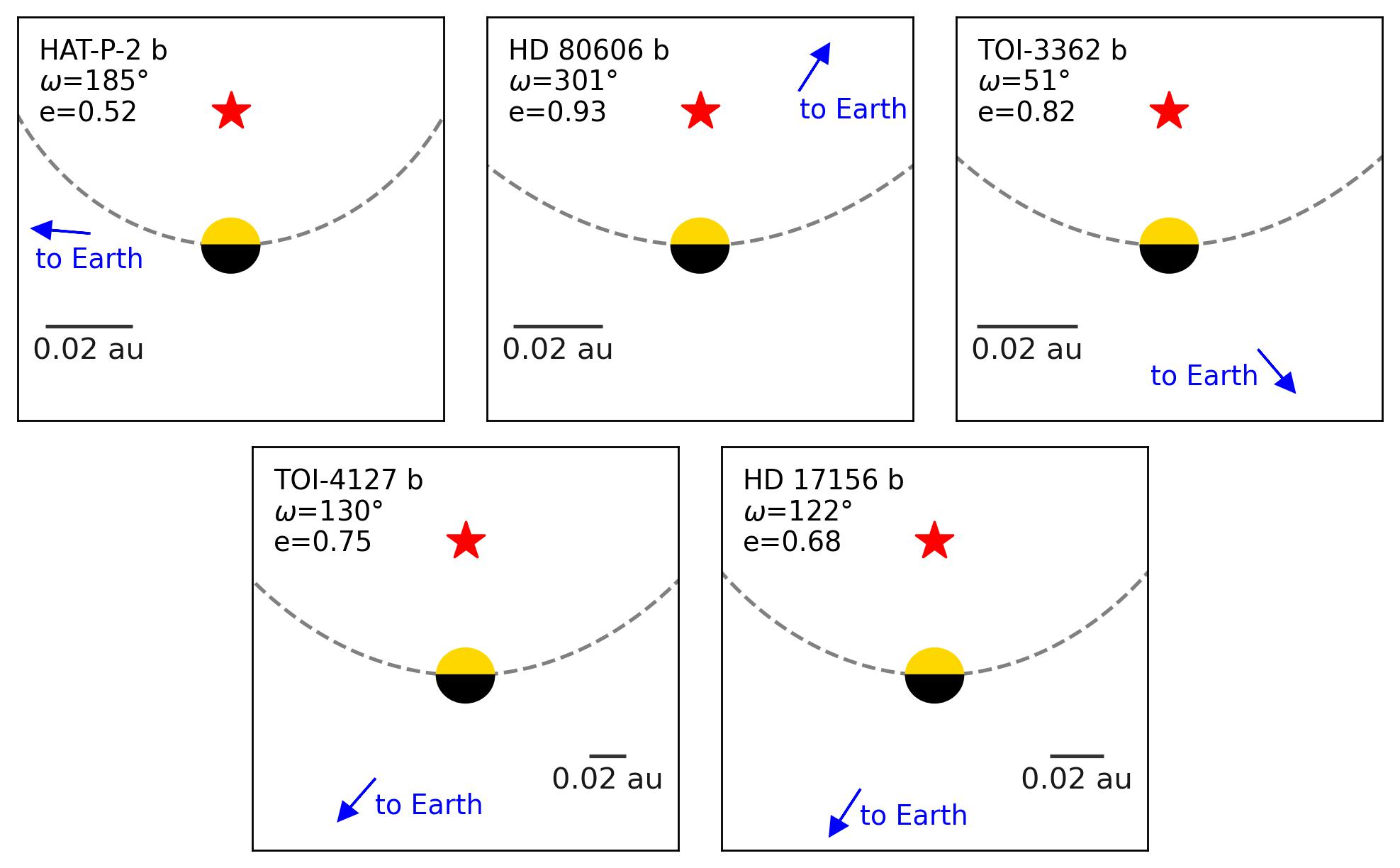}
    \caption{Drawings showing the orientation of the observer relative to the planet at its periastron passage for each of the five exoplanets (not to scale). The parent star is represented by a red star symbol and the planet by a disk, with the dayside coloured yellow and the nightside black. The blue arrow indicates the direction towards Earth. A solid black line shows the scale of the orbit. The orbital trajectory along which the planet moves in an anticlockwise direction is shown as a dashed grey line. The planet rotates anticlockwise as well.}
    \label{fig:visualisation orbits}
\end{figure*}

In this study we examined five exoplanets with a high eccentricity covering a range of orbital periods, including a number of targets that have been studied before and are to be revisited by JWST. 
The chosen planets are \mbox{\object{HAT-P-2 b}} \citep{Bakos2007}, \mbox{\object{HD 80606 b}} \citep{Naef2001}, \mbox{\object{TOI-3362 b}} \citep{Dong2021}, \mbox{\object{TOI-4127 b}} \citep{Gupta2023}, and \mbox{\object{HD 17156 b}} \citep{Fischer2007}. The relevant planetary, orbital and stellar parameters are listed in Table~\ref{tab:planet parameters}. The argument of periastron $\omega$ defines the viewing geometry, as visualised in Fig.~\ref{fig:visualisation orbits}. The rotation periods of the planets were computed using the commonly used pseudo-synchronous period $P_\textrm{ps}$ of \citet{Hut1981} \citep[e.g.][]{Kataria2013,Lewis2014,Lewis2017,Tsai2023a} given in Eq.~\ref{eq:pseudosync}, where the rotation period is synchronised with the orbital motion around the periastron passage since that is where the tidal effects are strongest. As a result, the substellar point moves little during the periastron passage: 
\begin{equation}
    P_\textrm{ps} = \frac{(1 + 3e^2 + 3/8e^4)(1-e^2)^{3/2}}{1 + 15/2e^2 + 45/8e^4 + 5/16e^6} P_\textrm{orb}
    \label{eq:pseudosync}
.\end{equation} 

Upon initialisation, the wind speed was zero everywhere. The temperature was horizontally homogeneous with a vertical profile comprised of three parts. The upper atmosphere at pressures smaller than 1 bar was given a constant temperature based on the analytical model of \citet{Parmentier2015}. That model requires an estimate of the intrinsic temperature, which is computed using the relation developed in \citet{Thorngren2019}. The layers deeper than 10 bar followed an adiabatic profile with a temperature of 1400~K at 1~bar, which \citet{SainsburyMartinez2019} find to be appropriate for the tidally locked hot Jupiter \mbox{HD 209458 b}. For the exoplanets modelled in this study, which have lower mean equilibrium temperatures than \mbox{HD 209458 b}, this deep temperature profile is likely too hot. However, \citet{SainsburyMartinez2019} advise using too hot a profile as atmospheric models converge more rapidly when cooling down than when heating up. Between 1 and 10~bar, the temperature profile is linearly interpolated in pressure. 

To reduce the total computation time, each planet was first treated as if it were tidally locked at a distance of the semi-major axis for 1000~d to let the temperature structure partly converge. Next, it was simulated for another 1000~d as an eccentric and non-synchronously rotating planet. On \mbox{\object{HD 80606 b}}, \mbox{\object{TOI-3362 b,}} and \mbox{\object{TOI-4127 b}}, the temperature structure or jet speed at the apastron still changed considerably between consecutive orbits, so the run continued for another 1000~d, after which the jet speed or the upper atmosphere's temperature did not monotonically change between orbits. However, the deep atmosphere's temperature was still not fully converged in any model. 
\citet{Schneider2022b} simulated the ultra hot Jupiter WASP-76~b for 86000~d and even after such a long run, the deep atmosphere's temperature had not fully converged. 
As such, achieving complete convergence is unfeasible. \citet{Mayne2017} and \citet{Carone2020} show that the dynamics of the deep atmosphere can considerably affect the upper atmosphere. However, \citet{Tsai2023a} indicate that additional internal heat from the overly hot deep atmosphere is unlikely to strongly affect the temperature and wind of the upper atmosphere of HD~80606~b, resulting in only a minor increase in planetary emission. Since the jet speed of the deep atmosphere has converged, we do not expect the incomplete temperature convergence of the deep atmosphere to notably affect our synthetic light curves. Therefore, our models are sufficiently converged for our purposes.

\subsection{Computing synthetic light curves}

Synthetic light curves were calculated during the post-processing of our GCMs using \mbox{\texttt{prt\_phasecurve}}\footnote{\url{https://prt-phasecurve.readthedocs.io/en/latest/}} \citep{Schneider2022a}, based on the \mbox{\texttt{petitRADTRANS}} code \citep{Molliere2019,Molliere2020}. We first computed a detailed radiation field throughout the atmosphere, which was then interpolated to a grid of 100 points on the side of the planet facing a hypothetical observer. We then integrated the outgoing flux over those 100 points and the chosen wavelength regime, making sure to take the viewing angle into account assuming an inclination of 90\degr. Finally, we divided by the stellar flux in the same wavelength range. For this work, we reduced the original forty streams used in the initial radiation field computation to six streams and the wavelength resolution from 1000 to 100. This drastically reduces the computation time of the light curve while affecting the precision of a test case based on the tidally locked exoplanet \mbox{HD 209458 b} by less than 1\%. 

We generated light curves in a wavelength range of 5~--~12~$\mu$m to represent the wavelength range of the Mid-InfraRed Instrument (MIRI) low resolution spectroscopy (LRS) mode aboard JWST. This instrument is commonly used to study exoplanets \citep{Rieke2015,Kendrew2015}, including some planets in our sample\footnote{Approved cycle 1 programme \#2008 {A Blast From the Past: A Spectroscopic look at the Flash Heating of HD80606b}. \url{https://www.stsci.edu/jwst/science-execution/program-information?id=2008}}. To gain some understanding of how sensitive the light curve is to the wavelength regime, we also separated the wavelength range into four sub-ranges, which approximately represent the F560W, F770W, F1000W, and F1130W imaging filters of MIRI \citep{Rieke2015}.

\section{Results}

\subsection{Pseudo-synchronously rotating models}

\begin{figure}
    \centering
    \includegraphics[width=88mm]{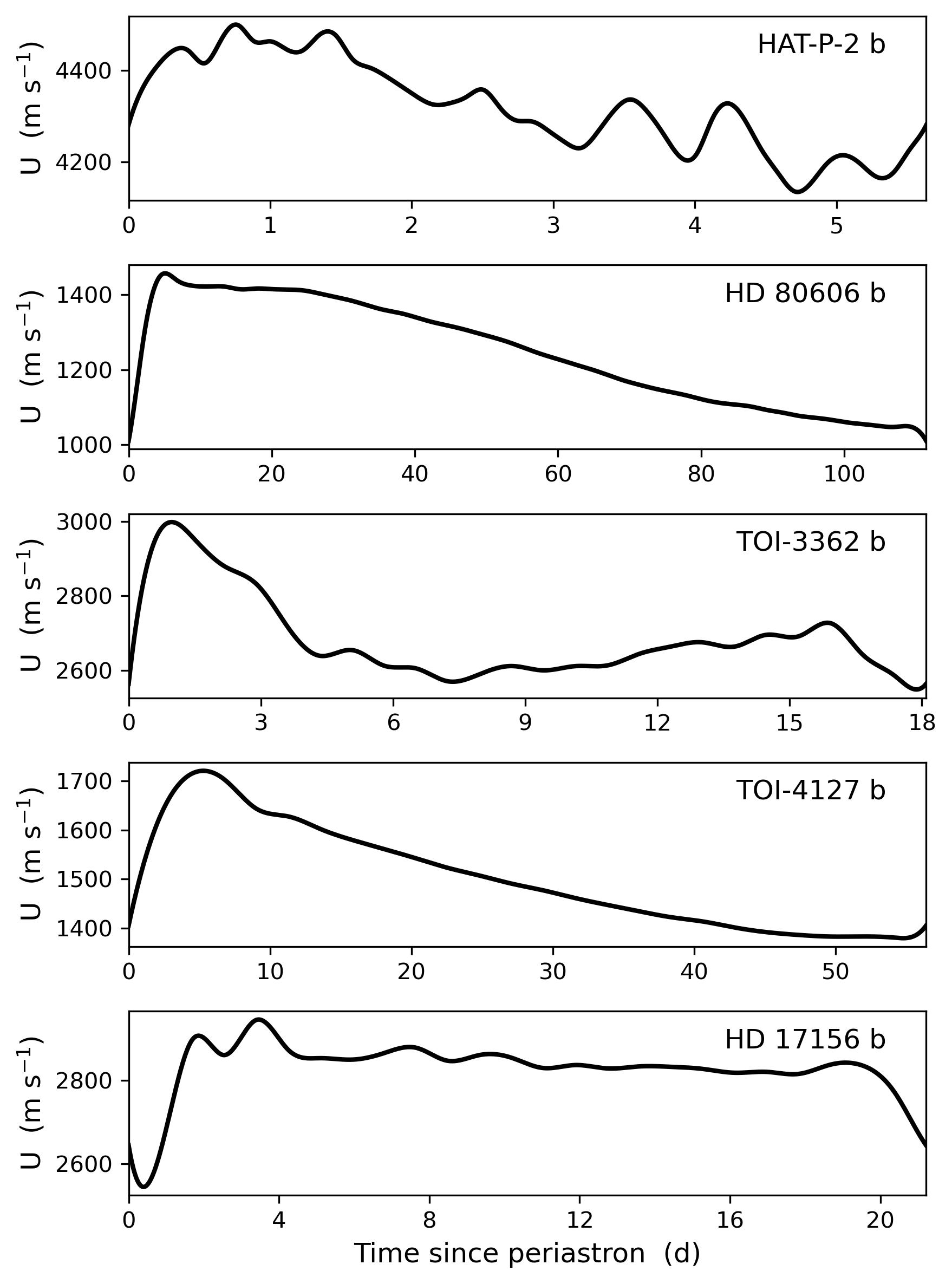}
    \caption{Speed of the equatorial jet throughout an orbit for each planet. The speed is computed as the maximum of the zonal wind speed between 100 and 0.01 bar after averaging over 1/50$^\textrm{th}$ of the orbital period, taking the zonal mean across the whole planet and the meridional mean between $\pm 20\degr$. We calculated the equatorial jet speed at 50 times and used cubic interpolation between these 50 times. Note that each panel has a different scaling on the two axes. The equatorial jet speed is quite consistent on \mbox{\object{HAT-P-2 b}}, \mbox{\object{TOI-3362 b},} and \mbox{\object{HD 17156 b}}, apart from some erratic behaviour around the periastron passage. On \mbox{\object{HD 80606 b}} and \mbox{\object{TOI-4127 b}}, however, it slows down considerably over the orbit after a period of acceleration after the periastron passage. }
    \label{fig:speedEvolution}
\end{figure}

\begin{figure*}
    \centering
    \includegraphics[width=162mm]{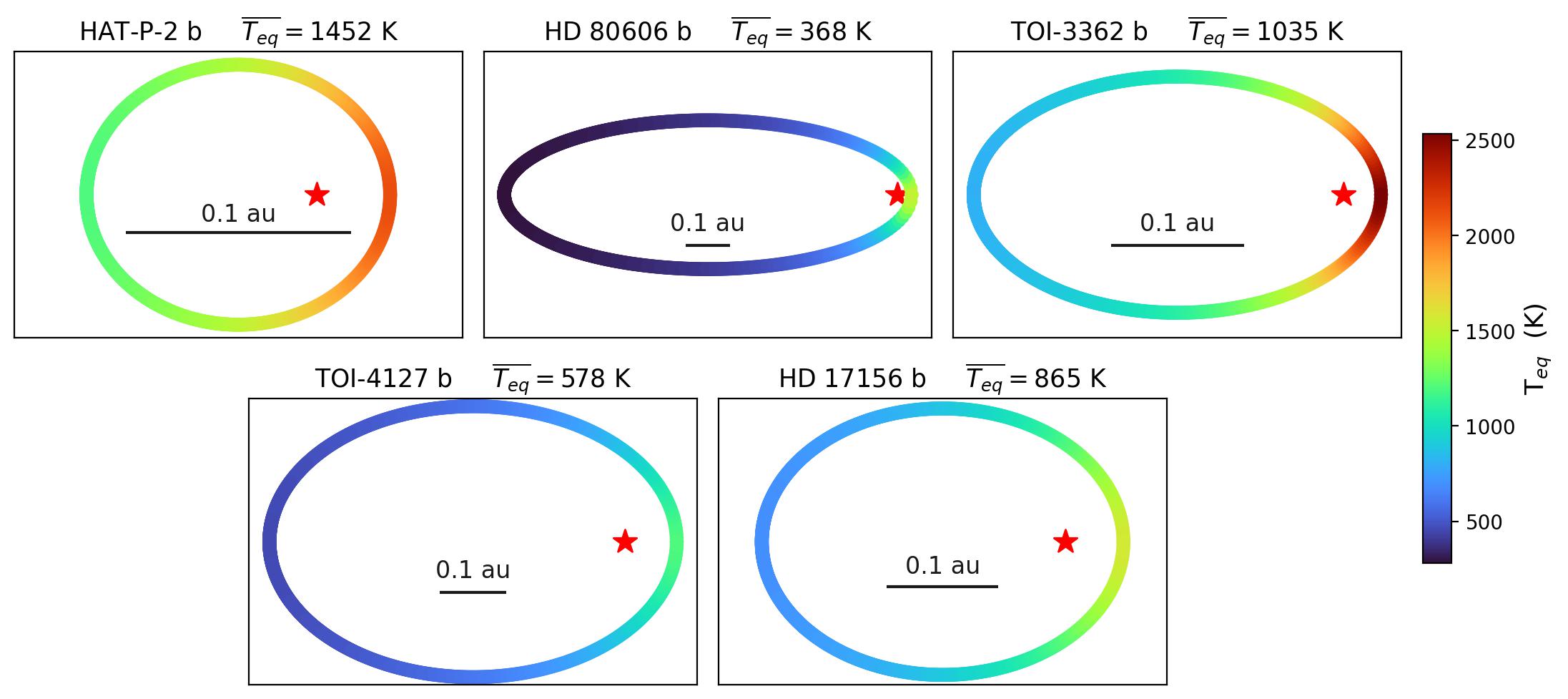}
    \caption{Equilibrium temperature assuming zero albedo over the orbit of the five exoplanets in our sample. A red star symbol indicates the position of the parent star. A solid black line shows the scale of the orbit. The effective temperature averaged over an orbital period is reported for each planet.}
    \label{fig:equilibrium temperatures}
\end{figure*}

Each of the five eccentric exoplanets in our sample features a prominent equatorial super-rotating jet throughout the entire orbit. \citet{Kataria2013}, \citet{Lewis2014,Lewis2017}, and \citet{Tsai2023a} find a jet on highly eccentric planets as well, demonstrating that the dynamical regime is quite robust. 
As in these previous works, the jet in the upper atmosphere is briefly replaced by fast day-to-night wind during the periastron passage, only to be quickly restored as the planet-star distance increases again. \citet{Lewis2017} and \citet{Tsai2023a} also note that the jet speed on \mbox{\object{HD 80606 b}} decreases over the orbit. Examining Fig.~\ref{fig:speedEvolution}, which shows the maximal jet speed of each planet over time, we see that the jet slows down from about \mbox{1400 m s$^{-1}$} to \mbox{1000 m s$^{-1}$} on \mbox{\object{HD 80606 b}}, in reasonable agreement with \citet{Lewis2017} and \citet{Tsai2023a}. \mbox{TOI-4127 b's} jet similarly slows down from \mbox{1700 m s$^{-1}$} to \mbox{1400 m s$^{-1}$}. The other three planets do not undergo such a consistent deceleration because their jets are continuously driven by the fairly strong irradiation they receive, as indicated by the equilibrium temperature shown in Fig.~\ref{fig:equilibrium temperatures}. The equilibrium temperature $T_\textrm{eq}$ is the black body temperature a planet needs to radiate away the irradiation. Neglecting the albedo effect, it is
\begin{equation}
    T_\textrm{eq} = T_\textrm{eff} \sqrt{\frac{R_*}{2r}}
.\end{equation}
Herein, $T_\textrm{eff}$ and $R_*$ are the effective temperature and radius of the host star and $r$ is the distance to the planet. As they are not as strongly irradiated, the thermal forcing on \mbox{\object{HD 80606 b}} and \mbox{\object{TOI-4127 b}} is insufficient to continuously feed the super-rotation. Instead, their jets are only driven around the periastron passage when the irradiation is intense and the substellar point does not move as much because of the assumed pseudo-synchronous rotation period, resulting in a thermal forcing regime that strongly resembles that of a tidally locked hot Jupiter. Outside the periastron passage, the jet continues under its inertia, but is slowly decelerated by friction processes. 

\begin{figure*}
    \centering
    \includegraphics[width=162mm]{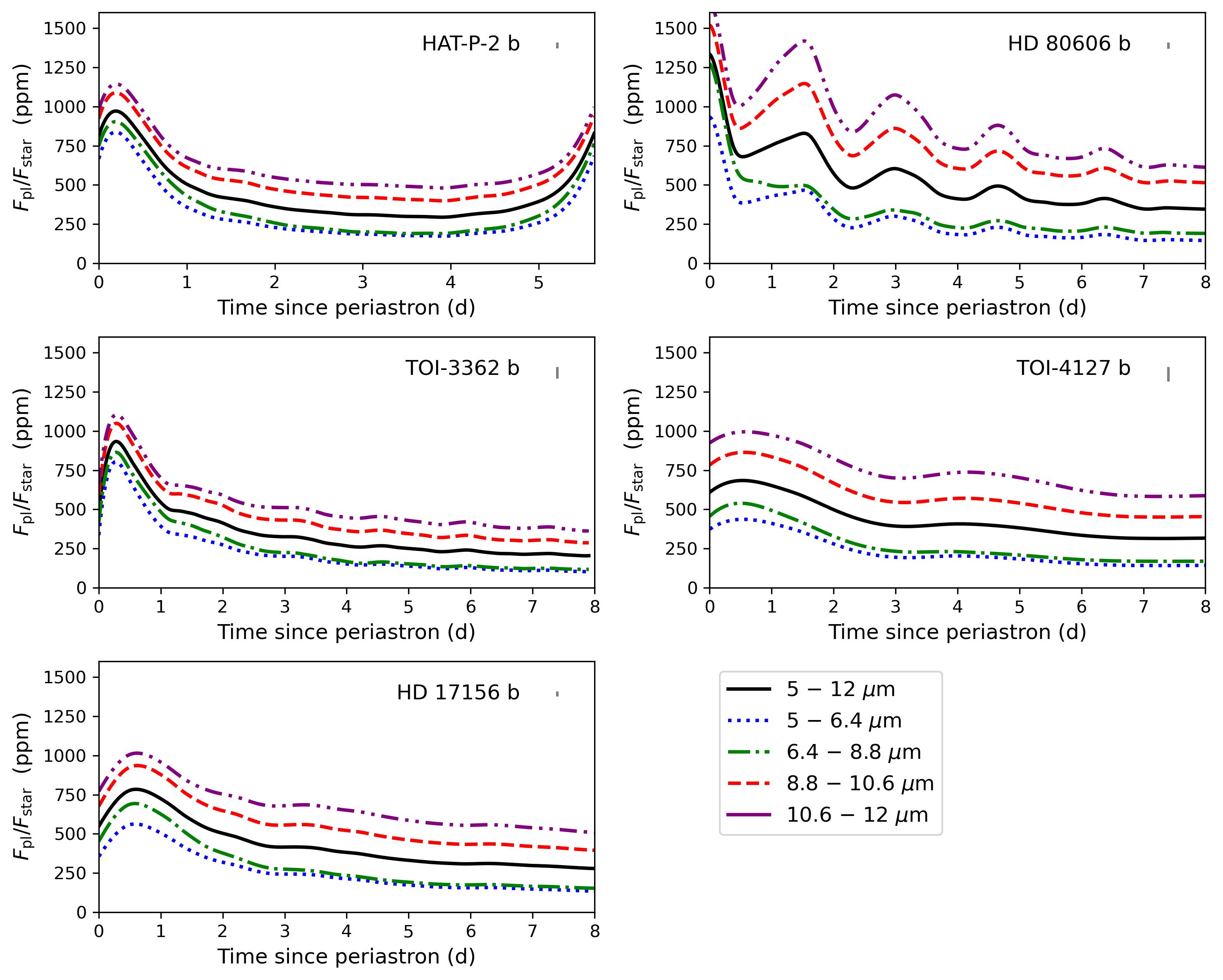}
    \caption{Synthetic light curves of all five planets relative to the brightness of the parent star in different wavelength ranges. We show the eight days after the periastron passage, except for \mbox{\object{HAT-P-2 b}} because of its short orbital period. A vertical grey line in the top right indicates the typical MIRI LRS precision when using the full 5 -- 12 $\mu$m range for a star of that visual magnitude. Ringing is seen as the local maxima in the light curves of \mbox{\object{HD 80606 b}} and \mbox{\object{TOI-4127 b}}, which is especially pronounced at longer wavelengths. \mbox{\object{HD 17156 b}} also displays a local maximum after the periastron peak in the long wavelength regimes. }
    \label{fig:all_lightcurves}
\end{figure*}

Figure~\ref{fig:all_lightcurves} displays the contribution of the planet to the total brightness of each system in different wavelength ranges. The solid black line shows the integral of the full 5~--~12~$\mu$m range. The planet's relative contribution is greater at longer wavelengths because of its cooler temperature, which is especially notable on \mbox{\object{HD 80606 b}}, \mbox{\object{TOI-4127 b,}} and \mbox{\object{HD 17156 b}}. Each planet's global maximum in emission occurs shortly after the periastron passage, except for \mbox{HD~80606 b's} due to its illuminated side being in the observer's view before the periastron passage, as shown in Fig.~\ref{fig:visualisation orbits}. After this initial periastron peak, the brightness of \mbox{\object{HAT-P-2 b}} and \mbox{\object{TOI-3362 b}} decreases mostly monotonically until the next periastron passage. \mbox{\object{TOI-4127 b}}, on the other hand, shows one local maximum after the first peak while \mbox{\object{HD 80606 b}} has four. These ringing peaks are especially high at longer wavelengths. \mbox{\object{HD 17156 b}} has local maxima with small amplitudes only at longer wavelengths.

\subsection{Breaking pseudo-synchronicity on HD 80606 b}

\begin{figure}
    \centering
    \includegraphics[width=88mm]{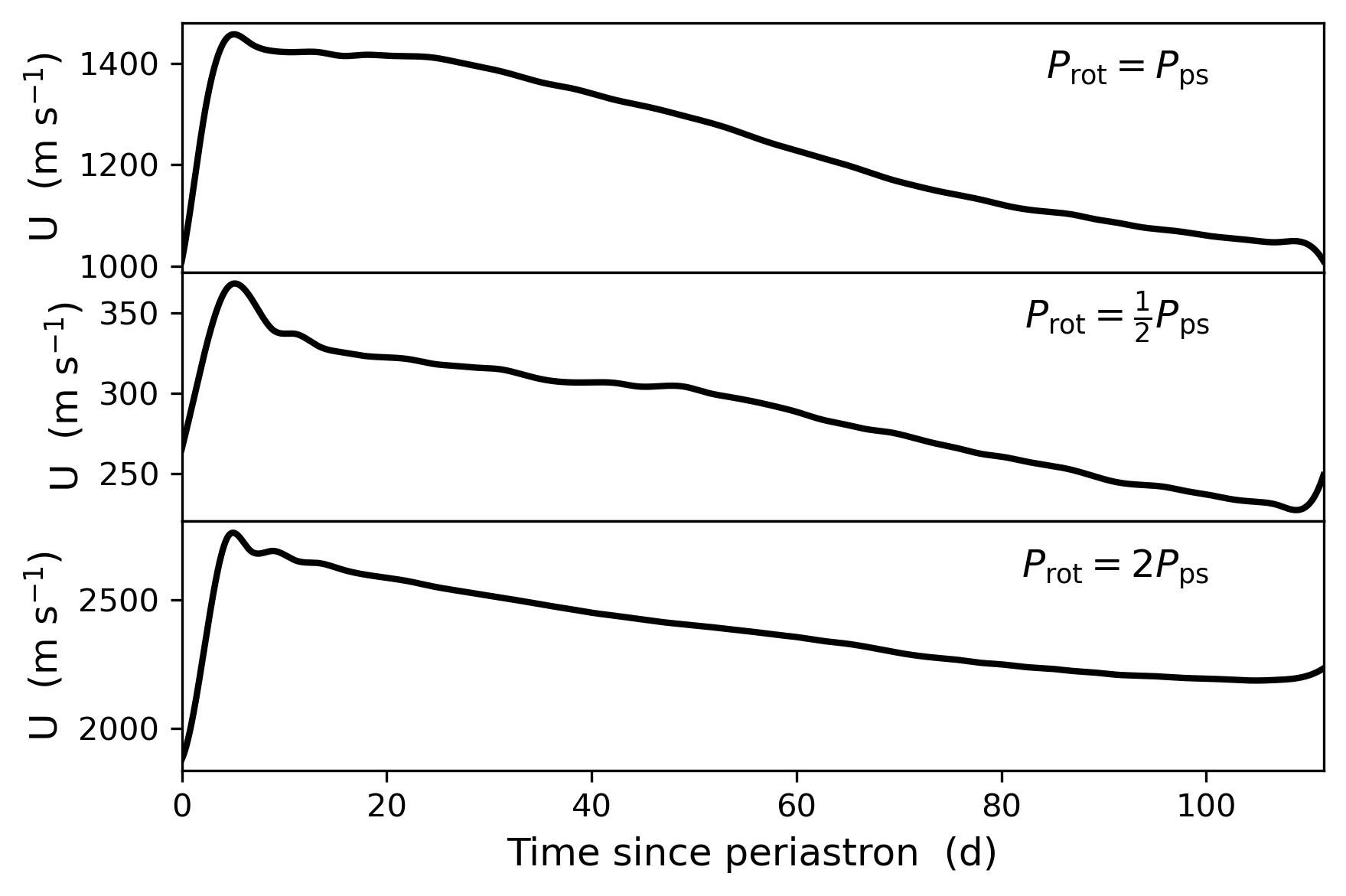}
    \caption{Speed of the equatorial jet on HD~80606 b with different rotation periods, computed using the method described in Fig.~\ref{fig:speedEvolution}. Each panel shares the same horizontal axis but has a unique vertical axis. The speed increases as the rotation period increases, although the jet is still only driven around periastron. When halving the rotation period, a consistent jet fails to form around the whole equator, so the zonally averaged zonal wind speed is decreased.}
    \label{fig:rotation_speedEvolution}
\end{figure}

To demonstrate the influence of the planet's rotation period, we repeated the \mbox{\object{HD 80606 b}} model with a rotation period equal to half and double the pseudo-synchronous period. Figure~\ref{fig:rotation_speedEvolution} shows how the speed of the equatorial jet differs from the pseudo-synchronous case. We note that a longer rotation period leads to a significantly faster equatorial jet. This is explained by the substellar point moving eastwards during the periastron passage if the rotation period is longer than pseudo-synchronous. This leads to more irradiation getting deposited into the already hot part of the atmosphere, thereby increasing the temperature gradient and thus the geostrophic wind speed, resulting in stronger angular momentum transport to the equator and hence a faster jet \citep{Showman2009,Showman2015}. Moreover, the jet is broader with a longer rotation period as explained by the equatorial Rossby deformation length $L_\textrm{D}$, a measure of the typical jet width given by
\begin{equation}
    L_\textrm{D, eq} = \sqrt{\frac{N k_\textrm{B} T R_\textrm{p} P_\textrm{rot}}{4 \pi \mu m_\mu g}}
    \label{eq:Rossby}
,\end{equation} 
with $N$ the Brunt-Väisälä frequency, $k_\textrm{B}$ Boltzmann's constant, $T$ the temperature, $\mu$ the mean molecular weight, $m_\mu$ the atomic mass, and $g$ the surface gravity \citep{Showman2002}. Consequently, the redistribution of heat becomes far more efficient with a longer rotation period. When decreasing the rotation period, the opposite occurs. The substellar point rapidly moves west at all times, even during the periastron passage, so the irradiation is spread thin and the resulting weak temperature gradient fails to pump sufficient angular momentum to the equator. 

\begin{figure}
    \centering
    \includegraphics[width=88mm]{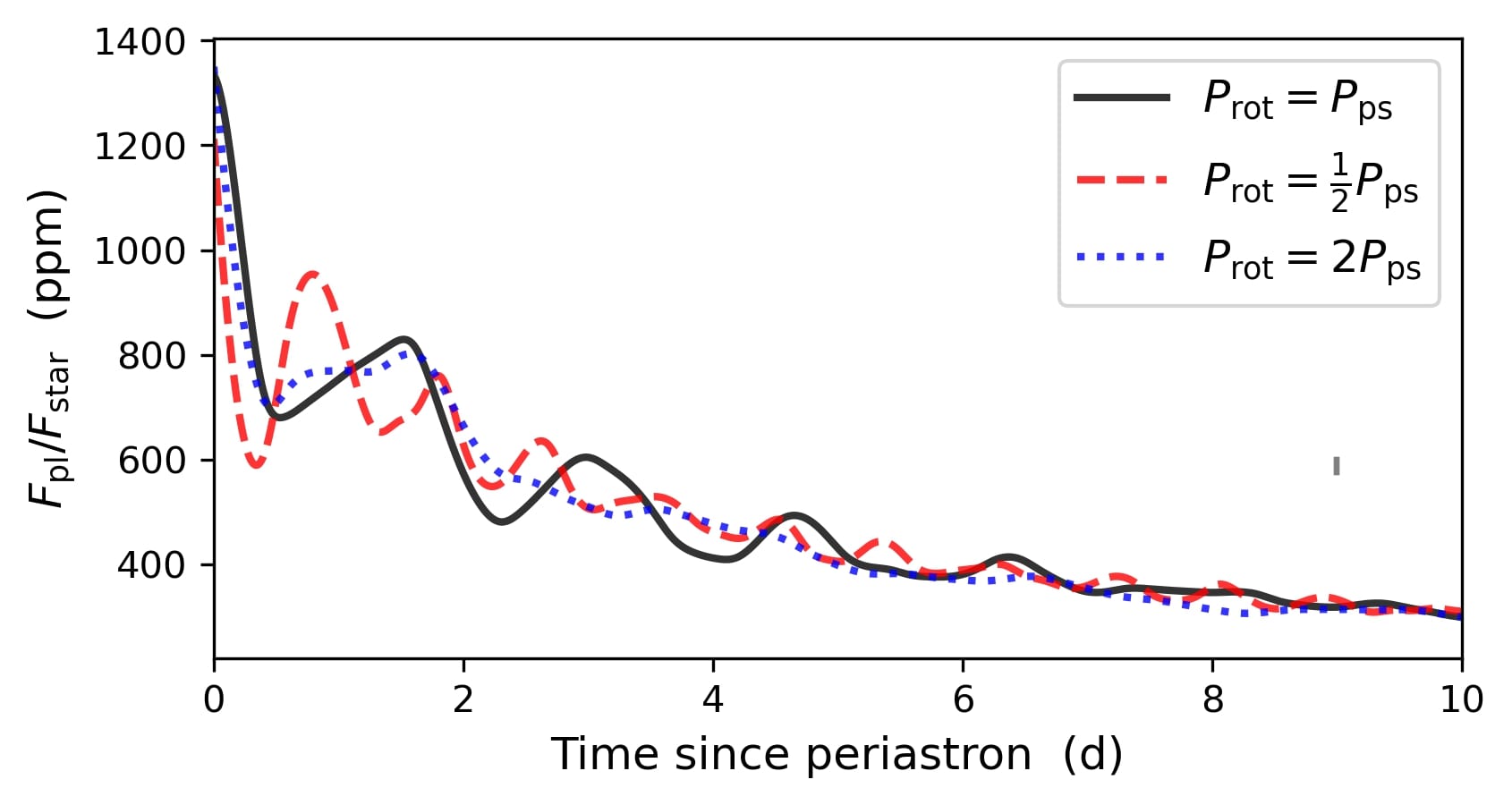}
    \caption{Synthetic light curves of \mbox{\object{HD 80606 b}} relative to the brightness of the home star for different rotation periods. The halved rotation period leads to more and stronger peaks forming, while the doubled period results in only one notable peak.}
    \label{fig:rotation_lightcurves}
\end{figure}

Figure~\ref{fig:rotation_lightcurves} demonstrates the importance of the planetary rotation period to the ringing phenomenon, as the case with a longer rotation period displays only one strong peak for two reasons. Firstly, heat is distributed more effectively by the circulation, which reduces the temperature contrast between the flash-heated and unheated hemispheres and hence the ringing amplitude. Secondly, as the rotation period is longer, it takes longer for the heated side to come into view, allowing it more time to cool and for the heat to be distributed. For the opposite reasons, the case with a short rotation period shows ten local maxima, the first of which are taller than in the pseudo-synchronously rotating case. Interestingly, the rapidly rotating case still shows small peaks over 8~d after the periastron passage, whereas the pseudo-synchronous case's last local maximum occurs about 6~d after the periastron passage, demonstrating that the lowered efficiency of heat redistribution by circulation plays an important role.

\section{Discussion}

\subsection{Timing of the ringing peaks}

Due to the strong increase in heating, the emission of the planet peaks sharply around the periastron passage on highly eccentric exoplanets. The timing of this peak relative to the periastron passage is affected by a number of parameters, including the planet's rotation period, thereby offering a rare opportunity for an observational constraint on planetary rotation, a route taken by \citet{deWit2016} and \citet{Lewis2017}. Consider as an example the emission of \mbox{\object{HAT-P-2 b}}, where about half of the dayside is in the viewing disk of an observer on Earth, as visualised in Fig.~\ref{fig:visualisation orbits}. In that case, one might expect the dayside to come entirely into view after $\frac{270\degr - \omega}{360\degr}P_\textrm{rot} \approx 10.7$~h. However, our synthetic light curve shown in Fig.~\ref{fig:all_lightcurves} predicts that the maximum in observed emission would occur $4.7$~h after the periastron passage. Here we ought to include the effect of the super-rotating equatorial jet, which pushes the gas heated during the heating flash eastwards, hence bringing it into view earlier. We can roughly estimate the jet-enhanced timing of the periastron peak to be 
\begin{equation}
    t_\textrm{peak} = \frac{270\degr - \omega}{360\degr} \frac{U_\textrm{rot}}{U_\textrm{rot} + U_\textrm{jet}} P_\textrm{rot} 
    \label{eq:jet enhanced timing}
,\end{equation}
where $U_\textrm{rot} = \frac{2 \pi R_\textrm{p}}{P_\textrm{rot}}$ is the eastward speed at the equator due to rotation and $U_\textrm{jet}$ the jet speed after periastron. For \mbox{\object{HAT-P-2 b}}, we estimate $U_\textrm{jet} = 4400$~m~s$^{-1}$ from Fig.~\ref{fig:speedEvolution} and $U_\textrm{rot} = 2500$~m~s$^{-1}$, resulting in $t_\textrm{peak} = 4.0$~h. This matches the predicted peak timing of $4.7$~h given the temporal resolution of about $2$~h.

The maximum in emission is also expedited by the cooling of the gas after the periastron passage as cooling leads to a decrease in emission over time. This is especially important shortly after the periastron passage, when the planet is still hot and the emission decreases more rapidly, as demonstrated by the $\propto T^{-3}$ dependence of the radiative cooling timescale $\tau_\textrm{th}$ given by
\begin{equation}
    \tau_\textrm{th} = \frac{P}{g} \frac{c_P}{4\sigma T^3}
    \label{eq:thermalTimescale}
,\end{equation} 
with $c_P$ the heat capacity at constant pressure and $\sigma$ the Stefan-Boltzmann constant \citep{Showman2002}. On \mbox{\object{TOI-4127 b}} for instance, we estimate the maximum to occur $21$~h after the periastron passage using Eq.~(\ref{eq:jet enhanced timing}), while the synthetic light curve predicts $13$~h after the periastron passage. As such, the cooling after periastron passage and the degeneracy between rotation and the super-rotating jet seriously complicate the extraction of the rotation period from the periastron peak, leading to large uncertainties in studies such as \citet{deWit2016}. 

\begin{figure*}
    \centering
    \includegraphics[width=162mm]{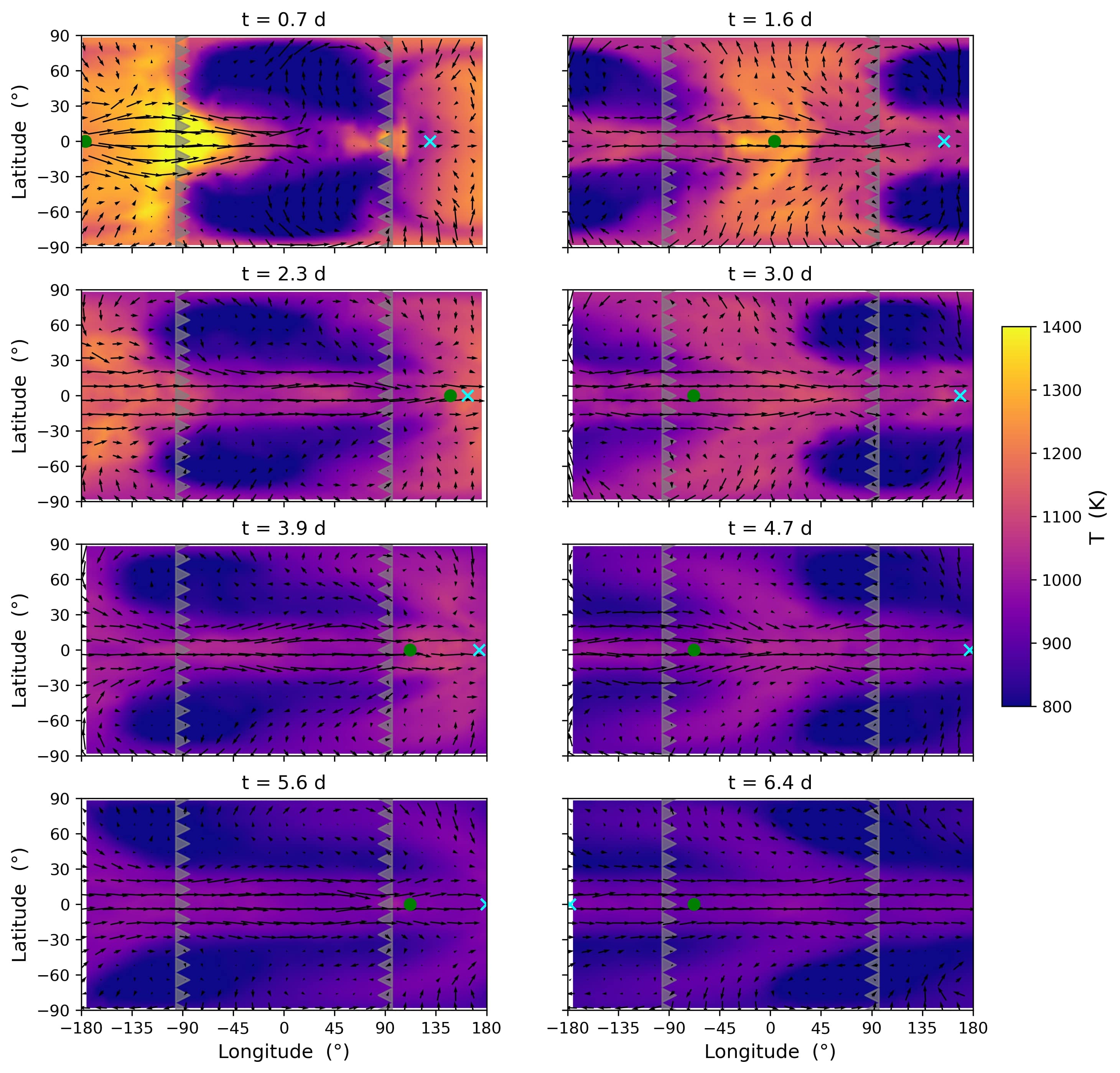}
    \caption{Horizontal slices of the temperature at 1 bar on \mbox{\object{HD 80606 b}} at different times after the periastron passage. The right panels correspond to observed emission maxima, and the left panels correspond to minima. Each panel is centred on the point below an observer on Earth, and grey triangles enclose the hemisphere in view, pointing towards the sub-observer point. Black arrows show the horizontal wind velocity. A cyan cross indicates the current substellar point and a green dot the substellar point at the periastron passage. }
    \label{fig:HD8 temperature change}
\end{figure*}

Under the right circumstances, ringing can isolate the rotation period from the timing of the ringing peaks. In the \mbox{\object{HD 80606 b}} simulation, the spacings between the peaks are $1.52$~d, $1.46$~d, $1.67$~d, and $1.72$~d with a resolution of about $0.05$~d, and the rotation period is $1.73$~d. The spacing between the periastron peak and first and second ringing peaks is shorter than the rotation period. To account for the super-rotating jet, we used
\begin{equation}
    \Delta t_\textrm{ring} = \frac{U_\textrm{rot}}{U_\textrm{rot} + U_\textrm{jet}} P_\textrm{rot} 
.\end{equation} 
Using $U_\textrm{rot} = 3000$~m~s$^{-1}$ and $U_\textrm{jet} = 1400$~m~s$^{-1}$ for \mbox{\object{HD 80606 b}}, we estimate $\Delta t_\textrm{ring} = 1.18$~d. This underestimates the modelled peak spacings in part because of the thinness of the jet due to \mbox{HD 80606 b's} short rotation period and low temperature reducing the equatorial Rossby deformation radius (see Eq.~(\ref{eq:Rossby})). Figure~\ref{fig:HD8 temperature change} shows the temperature evolution on \mbox{\object{HD 80606 b}} after periastron passage at 1 bar as seen by an observer on Earth. Initially, the flash heated gas is concentrated in one hemisphere. As the thin equatorial jet drags the hot gas at the equator eastwards, the spacing between ringing peaks is smaller than the rotation period. However, as the jet is quite thin, some flash-heated gas remains to the north-east and south-east of the substellar point at the periastron, indicated by the green dot. After an advection timescale given by 
\begin{equation}
    \tau_\textrm{ad} = \frac{2 \pi R_\textrm{p}}{U_\textrm{jet}}
    \label{eq:advection timescale},
\end{equation} the temperature on the equator is mostly homogenised by the jet. On \mbox{\object{HD 80606 b}} we estimate $\tau_\textrm{ad}=3.0$~d, matching the fourth panel of Fig.~\ref{fig:HD8 temperature change}. Yet the parcels of flash-heated gas outside the jet still maintain their higher temperature. As that gas continues to rotate in and out of view the last two ringing peaks follow with a spacing consistent with the rotation period. Consequently, the degeneracy between the rotation period and wind speed is broken, so the rotation period can be directly measured from the spacing between the last peaks.

\subsection{Detecting ringing on these five exoplanets}

Figure~\ref{fig:all_lightcurves} predicts that \mbox{\object{HD 80606 b}} and \mbox{\object{TOI-4127 b}} show some ringing in the MIRI LRS wavelength regime, while \mbox{\object{HD 17156 b}} would have some small local maxima at long wavelengths. For a solar-like star with an apparent magnitude of 9 such as HD 80606, MIRI is expected to reach a precision of approximately 25 ppm \citep{Greene2016,Batalha2019}. As such, we might expect at the four ringing peaks on \mbox{\object{HD 80606 b}} to be detectable. The first local maximum on \mbox{\object{TOI-4127 b}} may seem detectable as well, although its second, smaller bump would not be. However, considering its $V$-band magnitude is 11.44, the precision lies around 75 ppm instead, which the first ringing peak does not attain. Finally, the small local maximum at long wavelengths for \mbox{\object{HD 17156 b}} would not be detected, especially as one would have to limit oneself to the longer wavelengths of the instrument. As such, of the five exoplanets examined in this study, only \mbox{\object{HD 80606 b}} would exhibit any detectable ringing. 

\citet{Lewis2017} demonstrate that clouds may seriously diminish ringing on \mbox{\object{HD 80606 b}}. Since our model does not include the effects of clouds, we likely overestimate the amplitude of the ringing peaks. Similarly, \citet{Tsai2023a} show that including photo-chemistry in the post-processing of their \object{HD~80606~b} models slightly reduces the emission magnitude after the periastron passage. However, unlike clouds this is unlikely to qualitatively alter the light curve as the composition at photospheric altitudes returns close to pre-periastron levels by the time the first ringing peak occurs. 
On top of the lack of clouds, our model also assumes an obliquity of 0$\degr$. \mbox{\object{HD 80606 b}}, \mbox{\object{TOI-4127 b,}} and \mbox{\object{HD 17156 b}} have an orbital period greater than the 20~d limit beyond which one cannot safely assume tides to have aligned the rotational and orbital axes \citep{Rauscher2023}. Using a simple 2D shallow-water model, \citet{Ohno2019a, Ohno2019b} show how the obliquity and corresponding viewing geometry can strongly affect the light curve, which is confirmed by \citet{Rauscher2023}. If, for instance, one of the poles receives most of the flash-heating at the periastron, that heat would not be brought in and out of view by rotation, and hence no ringing. This too points towards our synthetic light curves overestimating the amplitude of the ringing. If little or no ringing is detected on \mbox{\object{HD 80606 b}}, this could be explained by a rotation period longer than pseudo-synchronous, clouds or obliquity. If clouds were responsible, one might expect to detect them using transit spectroscopy. If the non-detection of ringing were due to a long rotation period or an oblique orbit, the planet would prove itself useful to the research of tidal evolution and possibly to the detection of obliquity using secondary eclipse mapping \citep{Rauscher2017,Rauscher2023}.  

In our synthetic light curve of \mbox{\object{HD 80606 b}}, the final ringing peak takes place 6.5~d after the periastron passage. To include the decline after that local maximum, one would need a baseline stretching until approximately 7~d after the periastron passage. As the global maximum occurs shortly before the periastron passage, the baseline should be approximately 8~d long to cover all peaks. This baseline is considerably longer than those of currently accepted proposals for JWST observations of \mbox{\object{HD 80606 b,}} which were allotted a prime time of approximately one day\footnote{Approved cycle 1 programme \#2008, {A Blast From the Past: A Spectroscopic look at the Flash Heating of HD80606b}. \url{https://www.stsci.edu/jwst/science-execution/program-information?id=2008} and Approved cycle 1 program \#2488, {Real Time Exoplanet Meteorology: Direct Measurement of Cloud Dynamics on the High-Eccentricity Hot Jupiter HD80606 b}. \url{https://www.stsci.edu/jwst/science-execution/program-information?id=2488}}. As the amplitude of the ringing in our synthetic light curves is likely overestimated, it is uncertain whether the fourth ringing peak is indeed detectable. Therefore, it may be reasonable to shorten the baseline to omit the fourth ringing peak, in which case the baseline would stretch until approximately 5~d after the periastron passage. Alternatively, a weaker constraint can be placed on the rotation period through a comparison of observations to GCMs with different rotation periods as in Fig.~\ref{fig:rotation_lightcurves}. To clearly distinguish the $P_\textrm{rot}=P_\textrm{ps}$ case from the $P_\textrm{rot}=2P_\textrm{ps}$ case of \mbox{\object{HD 80606 b}}, a run ought to last for about 3~d after the periastron when the observed emission on the $P_\textrm{rot}=P_\textrm{ps}$ case increases while the $P_\textrm{rot}=2P_\textrm{ps}$ emission continues to decrease.

\subsection{Why only HD 80606 b displays ringing}

In our sample of five highly eccentric exoplanets, no planet displays any detectable ringing peaks except \mbox{\object{HD 80606 b,}} which has four. This remarkable difference in the level of ringing may be explained by a number of factors. Firstly, the extremely eccentric orbit of \mbox{\object{HD 80606 b}} results in a great difference in irradiation at the periastron compared to throughout most of the orbit, resulting in a large temperature contrast. As ringing is the consequence of a difference in emission between two hemispheres, a high eccentricity increases the amplitude of ringing peaks. Furthermore, under the assumption that the planet rotates pseudo-synchronously, a high eccentricity leads to a short rotation period, meaning little of the heat is distributed by circulation or irradiated away between rotations. Indeed, \mbox{\object{HD 80606 b}} has the second shortest pseudo-synchronous rotation period in our sample (see Table~\ref{tab:planet parameters}). However, \citet{Kataria2013} find an increase in the speed of the equatorial jet with eccentricity, which can raise the effectiveness of the circulation. Nevertheless, their work still suggests that a high eccentricity is beneficial for ringing, as they found no ringing on a hypothetical planet with $e=0.5$ and strong ringing on a similar case with $e=0.75$. 

Secondly, ringing in our sample is anti-correlated with the mean equilibrium temperature shown in Fig.~\ref{fig:equilibrium temperatures}. \mbox{\object{HD 80606 b}} has the lowest mean equilibrium temperature, while the two most irradiated cases, \mbox{\object{HAT-P-2 b}} and \mbox{\object{TOI-3362 b}} do not display any ringing. This is in part due to the long radiative cooling timescale of cool gas, which helps maintain the temperature contrast between the two hemispheres. Furthermore, weaker irradiation is known to lead to a slower equatorial jet and hence less efficient heat redistribution \citep{Kataria2013,Showman2015}. 
However, \citet{Kataria2013} note that a shorter orbital period and hence a greater equilibrium temperature leads to more and higher ringing peaks in their model with $e=0.75$. They indicate that the strengthened ringing is aided by the shortened pseudo-synchronous rotation period. As such, the net effect of the orbital period and mean equilibrium temperature is still unclear. 

Thirdly, \mbox{\object{HD 80606 b}} has the largest ratio of planetary radius to stellar radius of our five cases. This large ratio increases the planet's contribution to the total brightness of the system, hence increasing the ringing signal. Moreover, a large radius also means that circulation needs to travel longer to distribute heat across the entire planet, as demonstrated by the advection timescale in Eq.~\ref{eq:advection timescale}. Finally, the fairly low apparent visual magnitude of \mbox{\object{HD 80606 b}} means MIRI can achieve a high precision, making ringing easier to detect.

\subsection{More potential ringing targets}

\begin{figure}
    \centering
    \includegraphics[width=88mm]{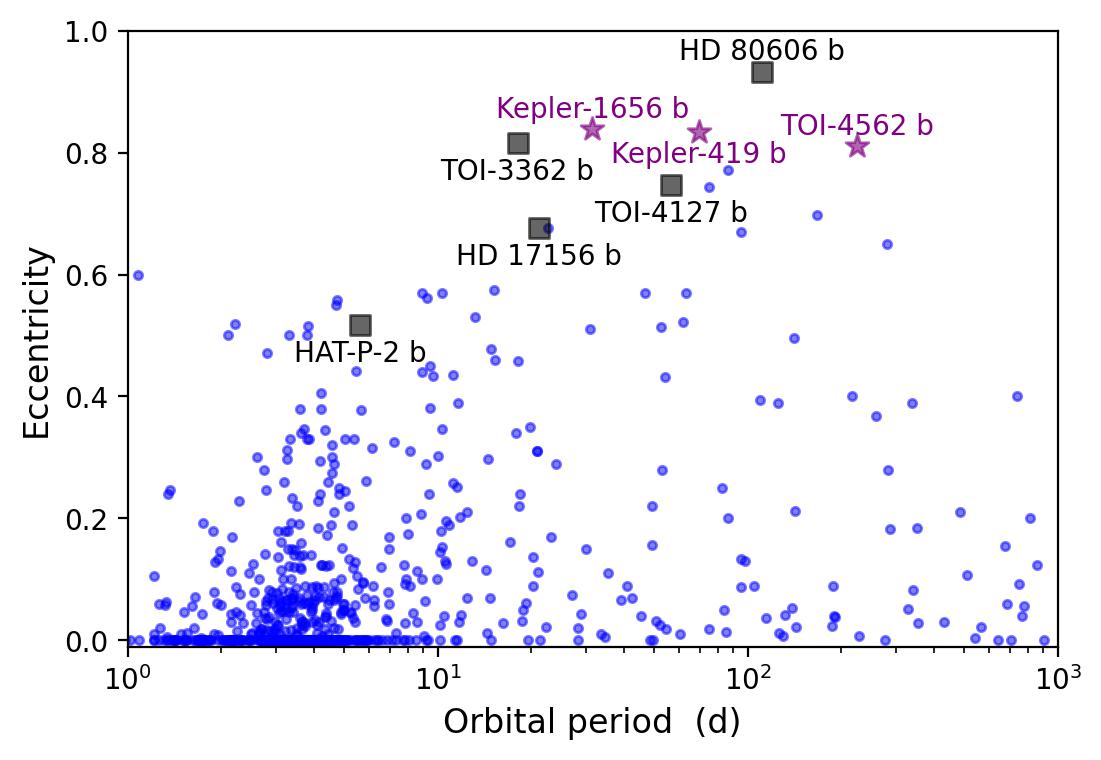}
    \caption{Orbital period and eccentricity of each confirmed Jupiter-like planet in the Encyclopaedia of Exoplanetary Systems \citep{Schneider2011}  as of 23 November 2023. Only planets with a mass of at least 0.1 $M_\textrm{J}$ and a radius of at least 0.4 $R_\textrm{J}$ are included. The black squares indicate the planets modelled in this work, and the purple stars mark three other potential candidates for the observation of ringing. }
    \label{fig:Porb_e_scatter}
\end{figure}

We conducted a search through the Encyclopaedia of Exoplanetary Systems\footnote{\url{https://exoplanet.eu/catalog/}} \citep{Schneider2011} for more ringing candidates with a high eccentricity, short pseudo-synchronous rotation period, large planet-to-star radius fraction and low visual magnitude. 
Figure~\ref{fig:Porb_e_scatter} shows the observed eccentricity and orbital period of each confirmed exoplanet in the encyclopaedia with an entry for these two quantities. Some exoplanets are not included in the graph because their eccentricity or orbital period is not given in the catalogue, even if they are available elsewhere in the literature as is the case for \mbox{\object{TOI-4127 b}} for example. In order to prioritise planets with well-constrained observed parameters, we limited ourselves to transiting systems. 

Some planets worth considering in future theoretical or observational studies of ringing are \mbox{Kepler-1656 b} \citep{Brady2018}, \mbox{TOI-4562 b} \citep{Heitzmann2023}, and \mbox{Kepler-419 b} \citep{Dawson2012}, highlighted by the purple stars in Fig.~\ref{fig:Porb_e_scatter}. Unfortunately, none of these planets seem as promising a target as \mbox{\object{HD 80606 b}}. \mbox{Kepler-1656 b} has a small radius of 0.45~$R_\textrm{J}$ and long pseudo-synchronous period while \mbox{Kepler-419 b} and \mbox{TOI-4562 b} orbit quite large and hot stars. Moreover, all three systems are rather dim with visual magnitudes of 11.67 for \mbox{Kepler-1656} \citep{Hog2000}, 12.09 for \mbox{TOI-4562} \citep{Zacharias2013}, and 13.04 for \mbox{Kepler-419} \citep{Zacharias2013}. Therefore, we recommend prioritising searches for ringing on \mbox{\object{HD 80606 b}}.

\section{Conclusions}

We examined which eccentric transiting exoplanets are likely to display detectable ringing, which is a series of detectable local maxima in the light curves due to the side of the planet that received the heating flash at periastron rotating into and out of view. To do so, we used the 3D GCM with full radiative transfer coupling \texttt{expeRT/MITgcm} and adapted it to include non-synchronous rotation and eccentric orbits. We simulated five highly eccentric transiting exoplanets, namely \mbox{\object{HAT-P-2 b}}, \mbox{\object{HD 80606 b}}, \mbox{\object{TOI-3362 b}}, \mbox{\object{TOI-4127 b,}} and \mbox{\object{HD 17156 b}}, using the pseudo-synchronous rotation period from \citet{Hut1981}. From these atmospheric models, we generated synthetic light curves in the wavelength range of the LRS mode of the MIRI aboard JWST. 

All five planets in our sample feature an equatorial jet throughout their entire orbit, showing that this dynamical regime is quite robust to changing irradiation due to eccentricity. However, the jets of \mbox{\object{HD 80606 b}} and \mbox{\object{TOI-4127 b}}, the planets with the longest orbital period in our sample, are only driven during the periastron passage and continue under their inertia while slowly decelerating. The planetary rotation period plays a critical role in the dynamical regime, which is exemplified well by the lack of a consistent equatorial jet when \mbox{\object{HD 80606 b}'s} rotation period is decreased to half pseudo-synchronous. The rotation period is also crucial to the ringing phenomenon, as it strongly affects both the amplitude and, especially, the number of ringing peaks. As a result, an observation of ringing or a lack thereof can provide a strong constraint on the rotation period, which is otherwise difficult to get. This would allow one to test the assumption of pseudo-synchronous rotation on highly eccentric planets. 

Our synthetic light curves predict local maxima on \mbox{\object{HD 80606 b}} and \mbox{\object{TOI-4127 b}}, while the other three planets only show some small humps in the 5~--~12~$\mu$m wavelength range. \mbox{\object{TOI-4127 b}'s} one ringing maximum is not detectable because of its rather dim parent star. For a 9 magnitude star such as \mbox{HD 80606}, JWST's detection threshold is roughly 25 ppm; this is attained by four ringing peaks on \mbox{\object{HD 80606 b}}, making it a promising observation target. However, our models neglect the effects of clouds and obliquity, both of which can obstruct ringing. Therefore, our synthetic light curves likely overestimate the amplitude of the ringing peaks. As a result, it is uncertain how many of the four predicted peaks on \mbox{\object{HD 80606 b}} are truly detectable. 
The timing of these peaks is affected by a complicated interplay between the planetary rotation period, equatorial jet speed, viewing geometry, and cooling after the periastron passage. If the ringing lasts longer than an advection timescale, however, the spacing between peaks approaches the rotation period, allowing one to extract the rotation period more reliably. An observation run with a baseline stretching seven days after the periastron passage would include all predicted peaks. A five-day-long observation would still cover three likely detectable peaks, enough to get the spacing between peaks after an advection timescale. 

From the study of these five exoplanets, we have tentatively deduced what factors facilitate observational ringing. They are a high eccentricity, a low apparent magnitude of the home star, a large planet-to-star radius ratio, and a short pseudo-synchronous rotation period. In our sample of five exoplanets, the ringing is more prominent on planets with a lower equilibrium temperature and a long orbital period; this contradicts the results of previous works, so the role of the orbital period is still unclear. Based on these conditions, we searched for other transiting exoplanets in the Encyclopaedia of Exoplanetary Systems that may display observable ringing but did not find a candidate as promising as \mbox{\object{HD 80606 b}}. Therefore, we conclude that \mbox{\object{HD 80606 b}} is currently the best known transiting exoplanet candidate for ringing.

\begin{acknowledgements}
The research presented in this paper was part of MV's master thesis, supervised by LD and co-supervised by RB, ADS and LC. 
ADS, LD and LC acknowledge funding from the European Union H2020-MSCA-ITN-2019 under Grant no. 860470 (CHAMELEON). LD further acknowledges support from the KU Leuven IDN/19/028 grant ESCHER. RB acknowledges support from the 'Origins' investment incentive. The authors would like to thank the unknown referee for their helpful comments. The data presented in this work were collected using the `pleiads' computer cluster of the Institute of Astronomy, KU Leuven, maintained by Maarten Dirickx. In this work, we have made use of the GCM post-processing library \texttt{gcm\_toolkit} \cite{Schneider2022a}.
\end{acknowledgements}

\bibliographystyle{aa}
\bibliography{ringing_paper_v3}

\end{document}